# Vital Node Identification in Complex Networks Using a Machine Learning–Based Approach

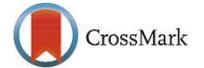

*Ahmad Asgharian Rezaei, Justin Munoz, Mahdi Jalili, Hamid Khayyam*

School of Engineering, RMIT University, Melbourne, Australia

A R T I C L E   I N F O

*Article history:*
Received 00 December 00
Received in revised form 00 January 00
Accepted 00 February 00

Keywords:
Influential Node Ranking
Complex Networks
Machine Learning
Vital Node Identification
Support Vector Machines
Influence Maximization
Epidemic Analysis

A B S T R A C T

Vital node identification is the problem of finding nodes of highest importance in complex networks. This problem has crucial applications in various contexts such as viral marketing or controlling the propagation of virus or rumours in real-world networks. Existing approaches for vital node identification mainly focus on capturing the importance of a node through a mathematical expression which directly relates structural properties of the node to its vitality. Although these heuristic approaches have achieved good performance in practice, they have weak adaptability, and their performance is limited to specific settings and certain dynamics. Inspired by the power of machine learning models for efficiently capturing different types of patterns and relations, we propose a machine learning-based, data driven approach for vital node identification. The main idea is to train the model with a small portion of the graph, say 0.5% of the nodes, and do the prediction on the rest of the nodes. The ground-truth vitality for the train data is computed by simulating the SIR diffusion method starting from the train nodes. We use collective feature engineering where each node in the network is represented by incorporating elements of its connectivity, degree and extended coreness. Several machine learning models are trained on the node representations, but the best results are achieved by a Support Vector Regression machine with RBF kernel. The empirical results confirms that the proposed model outperforms state-of-the-art models on a selection of datasets, while it also shows more adaptability to changes in the dynamics parameters.

## 1. Introduction

The rapid expansion of the Internet and social networks has given rise to new problems in network analysis such as identification of the most influential nodes and the dynamics of the spread of news and influence in complex and social networks [1]. A practical solution to the vital node identification problem has many applications in the domains of advertisement, training and news propagation [2]. Traditional approaches to this problem solely employ structural features such as the degree of the nodes [3], the length of the shortest paths between all pairs [4], the average length of shortest paths between all pairs [5], and the eigenvalue of the nodes [6] to rank the nodes based on their influentiality. While the adoption of structural features is widely popular, methods that accommodate these features generally suffer from high computational costs in addition to low accuracy. In another family of models, the idea of decomposing the network into different levels with high levels showing the inner parts of the network and low levels showing the outer parts, is used to rank the nodes assuming that nodes in the central parts of the network are more influential. This decomposition algorithm is also known as the K–Shell decomposition [7]. Inspired by this, many other methods including K-shell iteration factor [8], link significance method [9], extended K–Shell sum [10], and gravity centrality [11] adopted the idea and developed a customized K–Shell based method for ranking the influential nodes. However, this family of models fail to give accurate estimation of influentiality when underlying network is not so dense and has a low average degree.

* *Corresponding author.*
E-mail address: a.asghariyan.rezayi@gmail.com (A.A. Rezaei)

There are some other methods such as H–Index [12], and local H–index [13], local rank [14] and cluster rank [15] that mainly rely on the properties of their immediate neighbours for finding the vital nodes. These methods however are known to struggle with long influence paths, whereby many factors other than the immediate neighbours start to contribute to the actual influentiality of a node.

In another group of works a combination of K–shell decomposition and neighbourhood information is used to estimate the actual vitality. Mixed degree decomposition [16], cluster coefficient ranking [17], and mixed core, semi–local degree and weighted entropy [18] are some of these methods.

The main drawback of the above models is their inability to incorporate parameters of the dynamics in their ranking. In fact, these models always generate the same ranking of the nodes regardless of the parameters of the underlying dynamics. In an effort to resolve this shortcoming, Liu et al [19] used topological features along with dynamical properties to propose a dynamics–sensitive approach. Similar to this approach, Huang et al. [20] proposed a dynamics-sensitive method for influential node ranking in temporal graphs. In another research, Rezaei et al. [21] used randomized sampling of edges in a network for designing a dynamics–sensitive approach. There are also a number of works [22, 23] that rely on counting the number of paths and walks for designing dynamics-sensitive methods. However, the main drawback of the counting–based models is their high computational costs.

In contrast to the rule–based and heuristic models discussed above, recently, Zhao et al [24] proposed a data–driven machine learning-based method for identifying influential nodes in complex networks. They used a classification modelling of vital node identification problem, and trained their model on a big portion of nodes in the original network. In another research, Yu et al. [25] use network embedding in temporal networks for identifying the influential nodes. Similarly, Khajehnejad et al. [26] use adversarial graph embedding for finding the set of most influential nodes in networks with segregated communities. Hao et al. [27] also use network representation learning for finding influential nodes in networks with overlapping communities. The three latter works focus on specific versions of the vital node identification problem and do not apply to the general case. These methods also suffer from high computational complexity making them less applicable to large networks.

In this study we propose a novel and generic machine learning based approach for the vital node identification problem in its general form. Inspired by [28], our proposed method consists of three main steps: collective feature engineering, representative sampling, and training. In the collective feature engineering phase, elements of connectivity, degree and coreness of the nodes are put together to build a vector representation for each node. For having practical applications, a machine learning model for this problem should be trained on a small set of training samples. To this end, we propose a novel sampling method, called cluster sampling that ensures nodes of different structural and influential properties are included in the training set. We limit the sample size to be only 0.5% of the entire network, which leads to having small training sets, even for large networks. To deal with limited train data, we adopt a Support Vector Regression (SVR) machine with Radial Basis (RBF) kernel as our machine learning model. The empirical results show that the proposed method outperforms the existing state-of-the-art methods across different criteria while maintaining a high and stable performance even when the parameters of the underlying dynamic change, therefore showing a positive sensitivity to change in the dynamics.

## 2. Related Works

One can categorize the existing algorithms for influential node ranking into four groups of classical approaches such as neighbourhood-based method, K-shell based algorithms, Heuristic approaches, dynamics-sensitive approaches, and data-driven methods. In this section, we review some of the most significant works from each group.

To begin with, neighbourhood-based algorithms such as degree centrality [3] are among the easiest methods to explain. These methods attribute the vitality of nodes to the characteristics of their immediate or higher-level neighbours. For most of the methods in this family, the main advantage is low computational cost and effectiveness in cases when the length of the influence path is very short. H-index [12] is another method from this category of algorithms. For a node $u$, H-index is defined as the maximum value $h$ such that the degree of at least $h$ neighbours of $u$ is greater or equal than $h$. Local H-index [13] applies the neighbourhood philosophy in two-fold and leverages the H-index of neighbours of a node to calculate the LH score. If we show the H-index of nodes with function $H$, the following equation defines the LH score in terms of the H-index of itself and its neighbours.

$$LH(u) = H(u) + \sum_{v \in \Gamma(u)} H(v)$$

The idea of using the neighbourhood information twice or more forged other methods such as local rank [14] and cluster rank [15] and has inspired many other methods such as CNC+ [29], KS-IF [8], CRM [17], and DSR [30] as well. In the local rank method, the neighbourhood information is used in three steps. First the number of nearest neighbours and next nearest neighbours ($N(w)$) are counted for each node. Then the sum of this number for every node in the neighbourhood is calculated for each node again. And finally, the sum of sums is calculated as the final vitality of a node. In mathematical terms, we can write:

$$C_L(u) = \sum_{v \in \Gamma(u)} \sum_{w \in \Gamma(v)} N(w)$$

All the methods in the K-shell family use K-shell decomposition [7] as the basis of their method. For instance, the CNC score for a node $u$ is defined as the sum of the K-shell scores of its neighbours [29], or in mathematical terms:

$$CNC(u) = \sum_{v \in \Gamma(u)} KS(v)$$

In the same way, other algorithms of this family such as extended K-shell sum [10], K-shell iteration factor [8] and link significance [9] define a function of KS-score over the neighbours of a node as their final score. For instance, extended K-shell sum calculates the sum of K-shell scores in the neighbourhood of each node, and within 2-hops of a given node to approximate vitality. K-shell iteration factor [8] introduces a new parameter called iteration factor that differentiates between nodes in the same shell level. The final score is the sum of the K-shell based scores for nodes in the neighbourhood. In [9], link significance between two nodes $u, v$ is defined as $LS(u,v) = 1 - \frac{|\Gamma(u) \cap \Gamma(v)|}{|\Gamma(u) \cup \Gamma(v)|}$, and based on that the vitality of a node is defined as $I_u = \sum_{v \in \Gamma(u)} LS(u,v) K_s(v)$, where $K_s(v)$ is a function of the K-shell score for node $v$.

The heuristic approaches take one step further and try to combine the neighbourhood and coreness scores to come up with new algorithms. A good example of models in this family is and mixed core, semi-local degree and weighted entropy [18] that adds up a factor of degree, with a factor of coreness and entropy to approximate the vitality of the nodes. The following equation shows how MCDE score is calculated.

$$MCDE(u) = \alpha KS(u) + \beta|\Gamma(u)| + \gamma Entropy(u)$$

The classified neighbours [31] method is another algorithm from this family that uses K-shell algorithm to classify neighbours of a node into 4 different classes denoted as $e^u, e^{eu}, e^{el}$, and $e^l$ based on the K-shell score of the node and that of the neighbours. The vitality of nodes are then characterized by the following mathematical expression:

$$k_v^{CN} = \alpha \times |e_v^u| + \beta \times |e_v^{eu}| + \gamma \times |e_v^{el}| + \mu \times |e_v^l|$$

where $\alpha, \beta, \gamma$, and $\mu$ are tunable parameters and $e_v^u, e_v^{eu}, e_v^{el}$ and $e_v^l$ show the size of the nodes in each class.

In another algorithm of this family [32], authors employ a hierarchical approach and define two parameters, $f$, and $b$ that determine two aspect of the topological location of the nodes, closeness to the core of the graph, and distance from the periphery. The vitality of a node is then defined with a two-fold sum aggregating the influence spread across the immediate and second-order neighbours.

$$S(u) = \sum_{v \in \Gamma(u)} \sum_{w \in \Gamma(v)} |\Gamma(w)| \times (b_w + f_w)$$

In [17] authors use an improved cluster rank approach by incorporating common hierarchy of nodes and their neighbourhood information. In a rather different approach [30], authors tried to approximate the vitality of a node using the degree of diversity in its neighbourhood. Both methods perform an aggregation of the scores of their immediate and second-order neighbours to calculate the final influentiality of a node, but the chain of calculations in them is complex and cannot be summarized in this section.

The main drawback of the classical and K-shell based approaches is that they generate the same vitality scores no matter what the underlying dynamics and its parameters is. However, the influential ability of nodes changes if the dynamics change [33]. The first dynamics-sensitive approach [34] that tried to address this shortcoming assumed a linear approximation of non-linear dynamics such as Susceptible-Influenced-Removed (SIR) and defined vitality as the left eigen vector of the largest eigenvalue of the dynamics matrix. In a more intuitive approach [23], authors counted the number of possible walks of different lengths with the help of a decaying function that reflects the parameters of the underlying dynamics. Inspired by this work, another dynamics-sensitive approach for Susceptible-Infected-Susceptible (SIS) model was proposed that is and equivalent of path counting. Inspired by [34], authors of [20] assumed a linear approximation of diffusion propagation and proposed a differential function for capturing the vitality of nodes. In this regard, they used $x(t)$ to show the cumulative probabilities of nodes that are influenced between time 1 and $t$. Hence, $x(t) - x(t-1)$ will denote probabilities of the nodes that are influenced at time step $t$. For the first time step we set $x(1) = \beta A x(0)$, where $A$ is the adjacency matrix. Therefore, we can express $x(t) - x(t-1)$ in mathematical terms, and that would be:

$$x(t) - x(t-1) = \beta A [\beta A + (1-\mu)I]^{t-1} x(0)$$

In the above formulation, $I$ represents the identity matrix and $\mu$ is the recovery rate of the SIR model. If we set $\mu = 1$, and use $S(t)$ to show the cumulative spreading infleunce of all nodes up to time step $t$, then we have:

$$S(t) = (\beta A + \beta^2 A^2 + \cdots + \beta^t A^t)$$

However, since matrix multiplication for large networks is very costly, this approach is not practical for big graphs. In the most recent research [21], authors approximate influentiality of nodes through reachability of nodes in random sub-graphs sampled from the original graph. They prove in theory that if enough random-sub graphs are sampled from the original graph, the actual influentiality of nodes can be approximated with a factor of $1 - \epsilon$. Authors also show that in practice the proposed model outperforms the rest. The drawback of this method however, is its low adaptability to other dynamics. It is designed to work with the SIR dynamics, and the theories do not directly apply to other dynamics. In fact, this has been the main drawback of every model mentioned above, and the cure is machine learning based methods that can learn the rules themselves.

Only few works have recently started to solve influential node ranking with a machine learning based approach. In the closest research to our work [24], authors model the vital node identification as a classification problem where vitality of the nodes are shown by labels. They used 70% percent of the nodes in the network to train their model and used the remaining 30% for testing. Authors have also created labels of their own by grouping nodes based on their ground-truth vitality which is approximated through simulation. However, this approach proves to be impractical for large graphs as training a model on 70% of the network will have high computational cost. Moreover, using customized labels to represent vitality of the nodes, adds bias to the study. The bias comes from the fact that all the top ten nodes will receive the same label while it is important to know the order of vitality of nodes. There are also other methods from the family of machine learning based approaches that study a more specific version of the influential node ranking problem. In a recent work [25], critical node identification problem in temporal graphs is studied. Authors created an embedding of the nodes by attaching the adjacency matrix of the network for different snapshots. The machine learning model used in this work is a convolutional neural network with two layers. Authors showed that compared with the temporal version of baseline centralities, this model achieves a better performance. In another research [27], influential node ranking in overlapping communities is studied where authors attribute the most likely affiliation to nodes by maximizing the log-likelihood of having the existing connections in $G$.

In the results section we compare the performance of our proposed model against some of the most notable models mentioned above. We show that although existing models have advantages with respect to some performance metric, they perform poorly with respect to other metrics.

## 3. Proposed Method

Let us use $G = (V, E)$ to denote a complex network of nodes $V$ that are connected by a set of undirected edges, $E \subseteq V \times V$. Every two nodes that are connected by an edge are neighbours and we use the notation $\Gamma(v)$ to represent the set of neighbors of a given node $v$. The number of neighbours of a node is also referred to as the degree of that node. Since we are studying the problem of vital node identification in $G$, we first need to start with the definition of the vitality.

**Definition 1 (Vitality of a Node)**: In graph $G$, the vitality of a node is defined as the number of nodes that it can influence under a certain diffusion dynamic.

There are a number of diffusion dynamics that model the propagation of influence in complex networks [35]. In this work the main focus is on the Susceptible-Influenced-Removed (SIR) dynamics which has been widely used in the literature [17, 36]. The following table summarizes the propagation process in the SIR dynamics.

| Round | SIR Dynamics |
|---|---|
| 1 | Initially, there is only one node (say node $i$) in the Infected state and all other nodes are in the Susceptible state. |
| $t$ | All the nodes that are in the Infected state, can influence their Susceptible neighbours with probability $\beta$, and make them Infected. Infected nodes move to the Removed state with probability $\mu$. |
| end | The propagation finishes in the iteration in which there are no nodes in the Infected state. This means that the network has reached the steady-state and population of Susceptible and Removed nodes does not change. The number of Removed node is defined as the influence of node $i$. |

## 3.1 Collective Feature Engineering

The first step in any machine learning approach is feature engineering. In this work, inspired by graph representation learning research , we represent each node in graph $G$ with a collective feature vector that incorporates several elements including connectivity to other nodes, degree profile of the node (the node degree along with the degree of its neighbours), and node's coreness profile (coreness of the node together with the coreness of its neighbours) into the node's vector representation. In fact, the feature vector of a node is the Hadamard product of connectivity to the vector addition of degree and coreness vectors, each of which are defined formally hereunder:

**Definition 2 (Connectivity Vector, $C(v)$)**: The connectivity vector of a node $v$ is a vector of size $|V|$, where the entry at index $u$ is 1 if $(u, v) \in E$, otherwise it is 0.

**Definition 3 (Degree Vector, $D(G)$)**: The degree vector of a graph $G$ is a vector of size $|V|$ where the entry at index $u$ is the degree of node $u$ in $G$, $|\Gamma(u)|$.

Before defining the coreness vector, we need to define the notion of extended coreness score that we adopted for this work.

**Definition 4 (Extended Coreness Score)**: Assuming that $ks_u$ is the coreness score of node $u$ calculated through k-shell decomposition algorithm [7], the extended coreness score, $eks_u$ is calculated in the following way.

$$eks_v = ks_v \times |\Gamma(v)| + \sum_{u \in \Gamma(v)} ks_u \times |\Gamma(u)|$$

The introduced coreness measure helps to highlight the role of the neighbours of a node in the new coreness score. A node $v$ will have a high extended coreness score if $v$ and all its neighbours have high coreness scores. Between neighbours with the same coreness score, those with higher degree will have a higher contribution to the new coreness score. The Extended Coreness Score has similarity with the k-shell iteration factor [8], with the difference being that the iteration factor (which is either 1.5 or 2) is not included in the $eks$ formulation.

**Definition 5 (Coreness Vector, $K(G)$)**: The coreness score of a graph $G$ is a vector of size $|V|$ where the entry at index $u$ is the extended coreness score of node $u$, $eks_u$.

Given the definition for the degree, connectivity, and coreness vectors, we can now define the feature vector for each node $u \in V$ of the graph $G$.

**Definition 6 (Feature Vector, $X(u)$)**: The feature vector of node $u$ in a complex network $G$ is the Hadamard product of the connectivity to the vector addition of degree and coreness vectors, or:

$$X(u) = C(u) \circ ( \alpha_1 D(G) + \alpha_2 K(G))$$

where ∘ is the Hadamard product, + is vector addition and $\alpha_1, \alpha_2$ are tunable parameters.

## 3.2 Obtaining Representative Train Data

Having the engineered feature vector for each node, the next step is to create the training set. Since the ultimate objective is to train a model that can predict the vitality of a node given its feature vector, and as the actual vitality of nodes is unknown, we approximate vitality of nodes with multiple runs of the simulated underlying dynamics. In fact, we set the vitality of a node to be the average of the vitality of that node for multiple simulations of the underlying dynamics. Since running simulations of diffusion dynamics (such as SIR) is a computationally expensive task, we should limit the size of the training set to only a small portion of the nodes to minimize the pre-processing time. However, the training set should still be representative of all the nodes in the graph to help with an accurate training of the model. Leskovec and Faloutsos [37] have studied a number of sampling methods for graphs, including node-based sampling methods such as uniform node sampling and PageRank Sampling, edge-based sampling, and walked-based sampling approaches. Through their experimentation, node-based sampling methods were shown to outperform other sampling methods, particularly when graphs were induced through uniform node sampling. The authors also found that uniform node sampling can capture important properties of induced subgraphs better than any other sampling methods.

Inspired by the results presented in [37], we adopt uniform node sampling as our initial sampling method. Although it appears that a uniform node sampling method should be unbiased by its nature, when the input graph has a degree distribution similar to the power-law distribution, the uniform node sampling is in fact biased toward selecting nodes with low degrees. This fact is already highlighted in [38] as the inability of the uniform node sampling methods in retaining power-law degree distribution. To address this shortcoming, we propose a new sampling method, called cluster sampling, that aims to select nodes with different structural and influential properties. Similar to the other works in the literature [39], in this method, nodes are sampled from a feature space instead of the actual set of nodes. To obtain this feature space, we start by calculating the feature vector, $X(.)$, for every node in the network. For the sake simplicity of the model, we use the feature vector of the nodes as their representation. Further enhancement of the nodes' representation using auto-encoders [40] is postponed to future works. After obtaining the feature vector $X(.)$ for each node, we cluster the nodes in the $X$-space using a k-means clustering algorithm. To calculate the optimal value for the number of clusters, $k$, we try both the gap-statistics [41] and elbow methodss [42]. Eventually, for sampling $n$ nodes with this method, an equal number of $n/k$ nodes are uniformly sampled from each of the $k$ clusters. Since the feature vector, $X(.)$, already captures elements of node connectivity, degree and coreness, nodes within the same cluster share similar structural and influential properties. Therefore, sampling equal number of nodes from each cluster ensures that a representative variety of nodes are being selected.

Figure 1 shows a visual illustration of the cluster sampling method compared to the uniform node sampling. As it can be seen, in cluster sampling, sampled nodes are equally distributed among the clusters of the k-means algorithm (Figure 1-d), ensuring variety in the sampled set, while in the uniform node sampling method, each node is selected with the same probability. Therefore, in a network with power-law degree distribution, nodes with low degrees have a higher chance of being sampled by uniform node sampling method (Figure 1- b).

### 3.3 Modelling with Limited Training Data Obtained by Sampling

As discussed earlier, obtaining the training data for the vital node identification problem requires some pre-processing. It is because that the actual vitality of the nodes is unknown and needs to be approximated with the mean of vitality of the nodes in multiple simulation runs of the diffusion dynamics. However, this calculation is only required for the representative samples of the nodes, not the entire network.

To make the proposed method practical for large graphs, the size of the sampled nodes should be very small, otherwise obtaining the train data becomes the main bottle neck of this method. In this regard, we set the sampling size to be only 0.5% of the size of the network. This means for a network of 100,000 nodes, the training set will have about 500 sampled nodes of the network. Therefore, since the size of the training set is much smaller than the original network, we only limit our focus to machine learning models that can deal with limited/small training data. In this work we use Support Vector Regression (SVR) [43] which have been widely used in the literature for modelling data with small size [44]. The main idea behind the SVR is fitting a line to the training data that needs not to go through every data point, and can have up to a small deviation, call it $\epsilon$, from the training samples. We can formulate SVR as a convex program like the following:

$$\min \frac{1}{2}||w||^2$$
$$such\ that: |y_i - f(x_i, w)| \le \epsilon$$

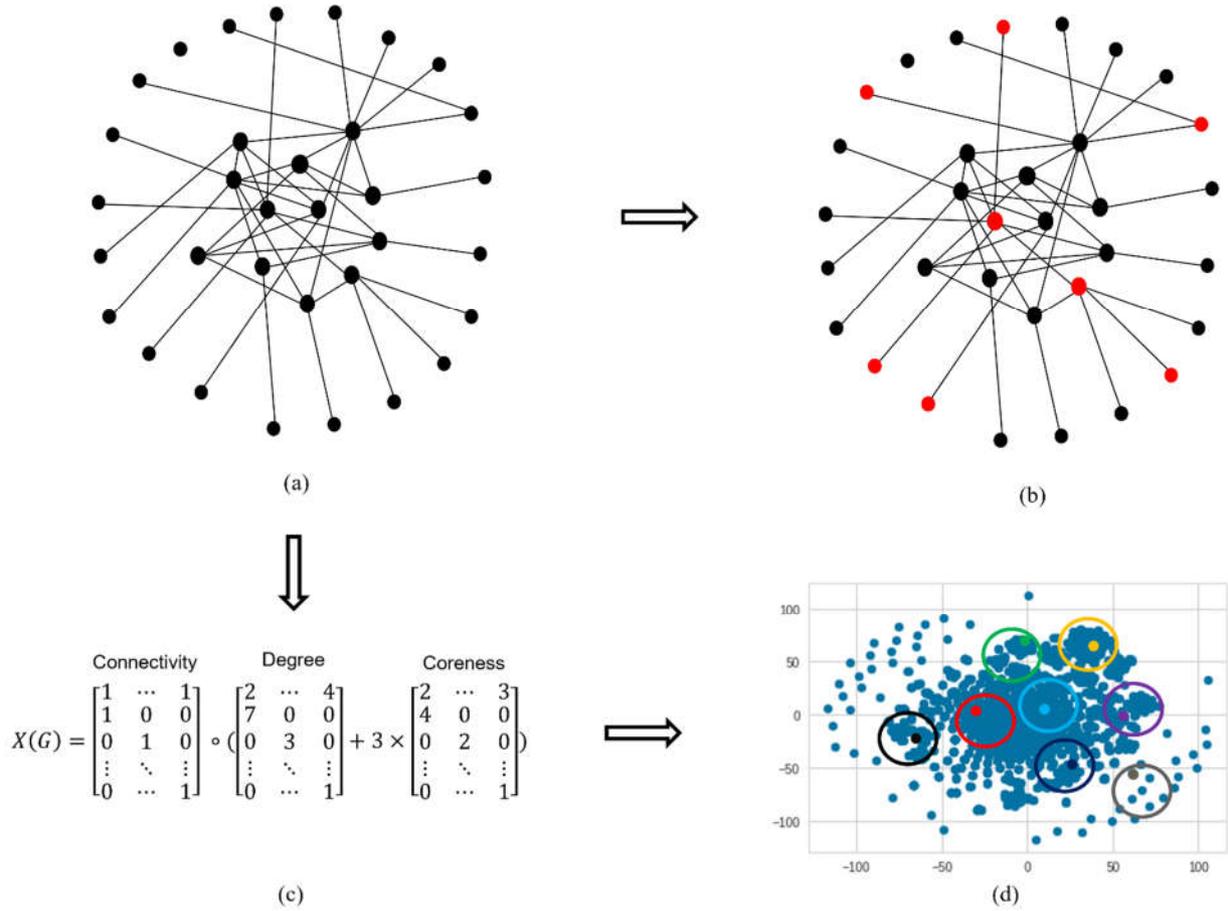

Figure 1. Uniform node sampling vs. cluster sampling. (a) An input network with power-law degree distribution. (b) Uniform node sampling. Sampled node are highlighted in red. Most of the sampled nodes have a low degree. (c) Building the feature matrix ($X(G)$) for the input graph. (d) Cluster sampling in the feature space ($X$-space). A 2-d representation of the feature space is shown in (1-d) where the coloured circles are clusters generated by the k-means algorithm. In cluster sampling, nodes are sampled from different clusters to ensure that nodes with different structural and influential properties are selected.

Assuming linear parametrization, the regression function will be $f(x, w) = wx + b$, and the model will be trained with the following loss function:

$$L(y, f(x, w)) = \begin{cases} 0, & |y - f(x, w)| \leq \epsilon \\ |y - f(x, w)| - \epsilon, & |y - f(x, w)| > \epsilon \end{cases}$$

This loss function implies that the model will incur a loss per data points that fall beyond the $\epsilon$ safe margin around the fitted line. Moreover, it also suggests that finding the best $w$ is independent of the number of training points, making SVR suitable for applications with limited train data. One of the limitations of a plain SVR model is its inability to capture non–linear relation between the independent variable $x$ and the target variable $y$. To address this issue, we use a kernelized version of SVR where a kernel transforms the data to a new space where the non–linearity in data is captured. The Radial Basis Function ($RBF$) [45] has been widely used as a kernel to SVR for capturing non–linearity. As shown below, RBF measures the similarity of a given point ($x$) to a kernel point ($x'$) using a radial distance and basis function. For a point $x$ which is close to the kernel point $x'$, $RBF$ will return a number close to 1, and for points that are far from $x'$ it returns a number close to 0. Mathematically we can say:

$$RBF(x, x') = \exp(-\frac{||x - x'||^2}{m\sigma^2})$$

where $m$ is usually set to be the number of features in the dataset. Assuming $k$ kernel points in the data, we can plug-in the $RBF$ formulation into the regression function of SVR, making it a kernelized SVR:

$$f(x, w) = \sum_{i=1}^{k} w_i \exp(\frac{||x - x_i||^2}{m\sigma^2}) + b$$

The SVR model with an RBF kernel can still be trained with the same loss function as of the loss for the standard SVR, $L(y, f(x, w))$. However, the SVR model alone will not generate the final vitality scores, and the final vitality score for a node, will be a function of the SVR predicted vitality for that node and its neighbours. In fact, we extend the machine generated scores by adding the SVR predicted vitality for the neighbours of a node to the final score of that node. As a result, we call this method Extended Machine Learning-based vital node identification or $EML$. Finally, for a given node $u$, we calculate the EML score, $EML(u)$, which reflects the vitality of that node using the following formula.

$$EML(u) = SVR(u) + \alpha \sum_{v \in \Gamma(u)} eks(v)\, SVR(v)$$

In the above equation, $SVR(.)$ denotes the predicted vitality by the kernelized SVR model, $\alpha$ is a tuneable parameter that takes values between 0 and 1, and $eks(.)$ is the extended coreness score defined earlier (definition 4).

The role of $\alpha$ in the above formulation to normalize the impact of the neighbours in the $EML$ score by assigning a weight lower than 1 to the predicted vitality of the neighbours. The $\Sigma$ term in the $EML$ score helps to differentiate between nodes based on their immediate neighbours and their vitality. With the help of this $\Sigma$ term, nodes that have more neighbours with high SVR predicted vitality with get a higher $EML$ score. Furthermore, in order to differentiate between the neighbours of a node $u$, we assign each node $v$ in the neighbourhood of $u$ a weight equal to its extended corness, $eks(v)$. This heuristic is based on the widely held assumption that the nodes in the central parts of the graph have higher vitality, and the assigned weight, $eks(v)$, magnifies the impact of these nodes in the $EML$ score. Algorithm 1 shows a pseudo code of the proposed method.

---

**Algorithm 1.** *Extended Machine Learning Based Vital Node Identification*
**Input**: An undirected, unweighted graph $G = (V, E)$, diffusion dynamics $df(.)$, and sample size $s$.
**Output**: A ranking of nodes in $G$ based on their vitality.
**Begin**
1:     $C(v) = \{1\ if\ (v,u) \in E,\ else\ 0\ \ \forall u \in V\}$     $\forall v \in V$     # Calculating the connectivity vector.
2:     $D(G) = \{\Gamma(v)\ for\ v \in V\}$     # Calculating the degree vector.
3:     $eks(v) = \{ks_v|\Gamma(v)| + \sum_{u \in \Gamma(v)} ks_u|\Gamma(u)|\}$     $\forall v \in V$     # Calculating the extended KS score for each node.
4:     $K(G) = \{eks(v)\ \ \forall v \in V\}$     # Calculating the coreness vector.
5:     $X(v) = C(v) \circ (\alpha_1 D(G) + \alpha_2 K(G))$     $\forall v \in V$     # Calculating the feature vector for each node.
6:     $k = opt(elbow(X(G)), gap(X(G)))$     # Obtain the optimum number of clusters, $k$.
7:     $clusters = Kmeans(\{X(v)\ \ \forall v \in V\}, k)$     # Cluster the nodes into $k$ clusters.

8:     $For\ i\ in\ (1, k)$:     # Pre-processing step for obtaining target vitalities for a small train set.
9:          $x\_sample = uniform\_without\_replacement(|s|/k, clusters[i])$
10:         $sampled\_x$.append$(x\_sample)$
11:         $sampled\_y$.append$(df(x\_sample))$     # $df$ is simulated to calculate the target vitality.
12:    $SVR = train\_SVR([sampled\_x, sampled\_y], kernel = \exp(-\frac{||x-x'||^2}{m\sigma^2}))$     # Training
13:    $For\ v \in V$:
14:         $EML(v) = SVR(v) + \alpha \sum_{u \in \Gamma(v)} eks(u)\, SVR(u)$     # Predicting vitality of nodes using the trained model.
15:    Return $\{(u_1, u_2, \ldots, u_n): EML(u_i) \geq EML(u_j)\ if\ i < j\}$
**End**

With having the $EML$ score computed for all the nodes, to obtain a ranking of nodes based on their vitality, the final output would a sorted list of nodes based on their $EML$ score.

In algorithm 1, line 6, $elbow()$, and $gap()$ are implementations of the elbow and gap-statistics methods for calculating the optimal size of clusters in the k-means algorithm. The $opt()$ function in line 6 selects a value for $k$ that can generate a final output with higher correlation with the ground truth. After building the clusters in the feature space (line 7), $|s|/k$ nodes are sampled per cluster (line 9) and added to the sample set along with the target vitality calculated at line 11. Lines 8-11 shows the cluster sampling process. For uniform node sampling (which is not shown in the pseudo codes), nodes are simply sampled with uniform probability from entire population of the nodes rather than clusters. The uniform node sampling is also done without replacement.

For an empirical analysis of the $EML$ algorithm, we test it on 8 different networks that are widely used in the literature. We then compare the experimental results of EML on these networks with 10 other baselines and state-of-the-art methods. The results show that EML outperforms the other models or achieves a competitive performance with respect to different criteria. A detailed discussion of the experimental results is included in the next section.

## 3. Experimental Results

Influential node ranking algorithms are usually evaluated through assessing their ability to generate a ranking of the nodes that has high correlation with the ground-truth ranking [32, 46]. For obtaining the ground-truth ranking, similar to what we have done in calculating the target vitality for the sampled nodes in Algorithm 1, the average of the vitality of the nodes in several runs of the simulated dynamics is set as the ground truth vitality of the nodes. The correlation of ranking highlights the ability of the model in estimating the vitality of nodes in general. To evaluate how good a model is in finding the top-k most influential nodes, other measures such as Jaccard Similarity is used [30]. However, the place of nodes in the generated ranking list is not the sole performance indicator for vital node identification models. There are other metrics such as the uniqueness of ranking and the distribution of ranks as well that are widely used for evaluation purposes [17, 31, 32].

In this study, we use a collection of 8 network datasets that have different structural properties. The datasets that we use in this study are Euroroad [47], Email, Hamster, Citeseer, Collaborations, Powergrid, Denaunay and PGP [48]. Table 1 details the specifications of the datasets used in this study. In this table $\beta$ is the influence probability in the SIR dynamics which is set slightly higher than the epidemic threshold, $\beta_{th}$, of the networks. Epidemic threshold is a value for the infection probability that can trigger a large scale infection [29]. The setting of the $\beta$ values in this study is in accordance with the suggestions made by Bae et al.[29], and is meant to enable large-scale influence spread. Since the we set the sample size to be only 0.5% of the network size, we selected networks that have at least a thousand of nodes so that the train set can have at least close to 10 samples.

Table 1. Details of the datasets used for the experimentation. Datasets with different size of nodes and different number of internal connections are selected.

| Dataset | Number of Nodes | Number of Edges | Max Degree | Average Degree | $\beta_{th}$ | $\beta$ |
|---|---|---|---|---|---|---|
| Euroroad | 1,174 | 1,417 | 10 | 2.414 | 0.333 | 0.35 |
| Email | 1,133 | 5,451 | 71 | 9.622 | 0.083 | 0.1 |
| Hamster | 2,426 | 16,631 | 273 | 13.711 | 0.024 | 0.03 |
| Citeseer | 3,264 | 4,536 | 99 | 2.779 | 0.274 | 0.3 |
| Collaborations | 4,158 | 13,422 | 81 | 6.455 | 0.0371 | 0.05 |
| PowerGrid | 4,941 | 6,594 | 19 | 2.669 | 0.258 | 0.3 |
| Delaunay | 8,192 | 24,547 | 12 | 5.992 | 0.073 | 0.08 |
| PGP | 10,680 | 24,316 | 205 | 4.554 | 0.053 | 0.1 |

## 3.1 Correlation of Rankings with Ground-Truth Around the Epidemic Threshold

Given the vitality of nodes identified by the algorithms, an important performance metric measures the correlation between the generated ranking and the actual ranking. Since the actual ranking is unknown, one way to obtain an approximation of that is to run numerical simulations of the underlying dynamics (SIR for instance). In this study, we conducted 3000 repeated numerical simulations of SIR to compute the ground-truth ranking. Nodes were ranked according to their influentiality, i.e., higher rankings were awarded to nodes with a higher influentiality. Throughout the numerical simulations, influentiality of the nodes is measured in terms of the number of nodes that they can influence, if the diffusion starts from them. To compute the correlation between the rankings generated by a given method ($R$) and the ground-truth ranking ($GTR$), we use the Kendall's tau correlation coefficient (KT-correlation) [49]. KT-correlation generates a value between −1 and +1, with a higher value showing a higher degree of correlation between the two rankings.

In this study, we use 10 existing methods to compare with the proposed method. The baselines and state-of-the-art methods are the k-shell decomposition centrality (KS) [7], mixed core and semi-local degree and weighted entropy (MCDE) [18], classified neighbours (CN) [31], k-shell iteration factor (Ks-IF) [8], dynamics-sensitive centrality (DS) [19], mixed degree decomposition (MMD) [16], H-index method (H-Index) [50], diversity-strength ranking (DSR) [30], link significance (LS) [9] and entropy based ranking measure (ERM) [17]. Table 1 summarises the KT-correlation of these methods with the ground-truth ranking over different networks. In this table Column EML-UNode shows the correlation results for the EML node with uniform node sampling method, while EML-Cluster summarizes the results of the EML method with the cluster sampling method. As highlighted in Table 1, the two versions of EML, outperform the other methods over six networks, while obtaining a close to best performance over the other two networks (Email and Delaunay). When compared head-to-head, both versions of the EML method achieved competitive performances. EML-Cluster was shown to be superior across four datasets, while EML-UNode across three datasets. Although it appears that EML-Cluster has better performance just as expected in the theory, however, the performance of EML-Cluster is surprisingly high and exceed what we expected in theory. The results show that even for a network with power-law degree distribution (the Citeseer network), the uniform sampling method outperforms cluster sampling. In this case, although nodes with low degree have higher chance of being sampled, however, the feature vector is probably informative enough to help the model with learning the dependency relation between node's degree, node's feature vector and the target vitality, even when most of the nodes have low degrees. Therefore, in the rest of this paper, we only report the results of EML-UNode which is denoted by EML.

## 3.2 The Impact of the Influence Probability, $\beta$ of SIR Model, on the Correlation of Rankings

In this section we study the impact of changing the influence probability, $\beta$ of SIR model, on the correlation of rankings with the ground-truth ranking. Previously, we observed that the rankings generated by the EML models outperforms the other models if the influence probability is close to the epidemic threshold. We repeat the same experiment for several times, each time with a different influence probability. Figure 2 shows the results of this experiment over four different networks of different sizes and connections density.

| Datasets | KS | MCDE | CN | KS-IF | DS | MMD | H-Index | DSR | LS | ERM | EML-UNode | EML-Cluster |
|---|---|---|---|---|---|---|---|---|---|---|---|---|
| Euroroad | 0.6034 | 0.639 | 0.6087 | 0.8013 | 0.85 | 0.597 | 0.6063 | 0.7946 | 0.6933 | 0.8646 | 0.8934 | **0.8965** |
| Email | 0.8717 | 0.8478 | 0.8478 | 0.892 | **0.9312** | 0.8013 | 0.8737 | 0.8911 | 0.9184 | 0.9011 | 0.9032 | 0.9032 |
| Hamster | 0.7158 | 0.7243 | 0.7307 | 0.8748 | 0.8042 | 0.7032 | 0.7189 | 0.8047 | 0.8767 | 0.8339 | **0.8941** | 0.8925 |
| Citeseer | 0.561 | 0.5522 | 0.5946 | 0.7392 | 0.7803 | 0.5254 | 0.5578 | 0.7376 | 0.6841 | 0.7418 | **0.8266** | 0.8177 |
| Collaborations | 0.7131 | 0.7349 | 0.7313 | 0.7815 | 0.5703 | 0.722 | 0.7207 | 0.77 | 0.7392 | 0.7735 | 0.7701 | **0.7878** |
| PowerGrid | 0.5486 | 0.5869 | 0.5507 | 0.7668 | 0.8268 | 0.5494 | 0.5571 | 0.7262 | 0.5963 | 0.774 | 0.8423 | **0.8490** |
| Delaunay | 0.492 | 0.7744 | 0.5138 | 0.6913 | 0.4837 | **0.7906** | 0.5483 | 0.1663 | **0.7907** | 0.7642 | 0.7800 | 0.7821 |
| PGP | 0.4918 | 0.4916 | 0.5083 | 0.7175 | 0.6942 | 0.4776 | 0.4925 | 0.7199 | 0.6882 | 0.7158 | **0.7288** | 0.7282 |

Table 2. Correlation of rankings generated by different models with the ground-truth ranking generated through simulation.

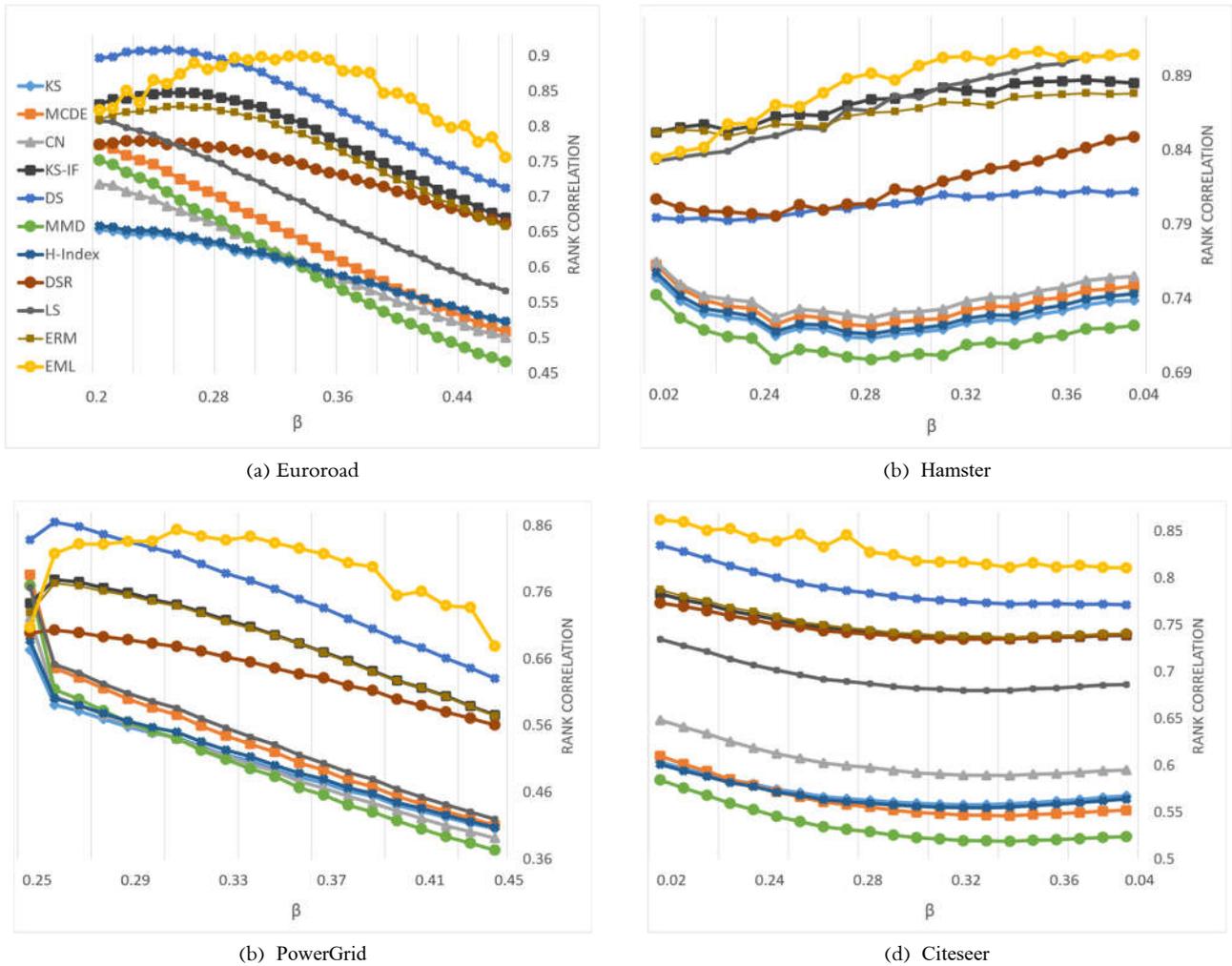

Figure 2. The impact of changing the dynamics parameters on the correlation of rankings with ground truth for four different datasets. The proposed model shows the most stable behaviour when the influence probability changes.

The diagrams show the ranking correlation with ground–truth as a function of $\beta$. In three of the networks, the EML model is outperformed over small values of the influence probability, $\beta$, which can be attributed to the fact that when $\beta$ is low, only few nodes are influenced in the diffusion process. Therefore, the target vitality for most of the nodes will be zero or close to zero. This leaves less information for the SVR model to learn from resulting in poor performance. Other than that, the performance of the EML model shows the best stability when $\beta$ increases. The other method that is also showing good stability to the change of the influence probability, is the dynamics–sensitive centrality (DS) method [19]. As discussed earlier, this method performs matrix multiplication to simulate the diffusion process. The main drawback of this method is that it tries to approximate a potentially non-linear diffusion dynamic with linear operation. We can see that the performance of this model is very dependent on the density of the connections. In fact, when connections are not so dense, such as Euroroad, PowerGrid and Citeseer, the model is showing reasonable performance. However, for dense networks, such as Hamster, the model shows a mediocre performance. The reason behind this is that, for low–density networks, the diffusion process is simpler with fewer steps and can be approximated with a linear function, while for dense networks, the diffusion becomes highly non–linear and linear approximation will be inaccurate. The main takeaway from the comparison between EML and DS is that, although DS is designed to be sensitive to any change in the dynamic, our proposed method, EML, shows even a better sensitivity to the SIR dynamic. This is probably because, unlike any other model, EML is trained with ground-truth vitality, and since any change in the dynamics impacts the ground-truth vitality, any change in the dynamics is also reflected by the EML model too.

Other than EML and DS that can handle changes in the dynamics, the other methods are dynamics-independent and generate the same ranking always. That is why we see a decline in the performance of these methods as $\beta$ changes. Among these methods, those that solely rely on immediate neighbours for estimating nodes vitality, such as CN, MMD, H-Index and MCDE perform poorly, especially for high values of $\beta$. One can attribute this low performance to the big length of the influence paths in the test networks which implies many other factors, such as nodes in the middle of the path contribute to the existence of the path, while such factors are neglected by these models.

### 3.3 The Ability to Find the Top-K Most Influential Nodes

Although the correlation of rankings with ground-truth is an indicator of the overall ability of a model in identifying the influentiality of nodes, it tells nothing about the ability or accuracy of a model in identifying the most influential nodes. To this end, in this section, we use a new metric in which the set of top-k most influential nodes identified by a given model is compared against the top-k most influential nodes from the ground-truth. Similar to section 3.1, the ground-truth is computed empirically. Many metrics such as precision or recall can be used to show the degree of similarity between the top-k ranks of models and ground-truth. Here we use Jaccard similarity [51] as it is more sensitive to mismatches between two lists. Jaccard similarity coefficient, $J(k)$, for the top-$k$ elements of two lists, $R$, and $GTR$ is mathematically defined as $J(k) = \frac{|R(k) \cap GTR(k)|}{|R(k) \cup GRT(k)|}$. Let us assume that in the above formulation, $R$ is the ranking generated by a given model, and $GTR$ is the ground-truth ranking.

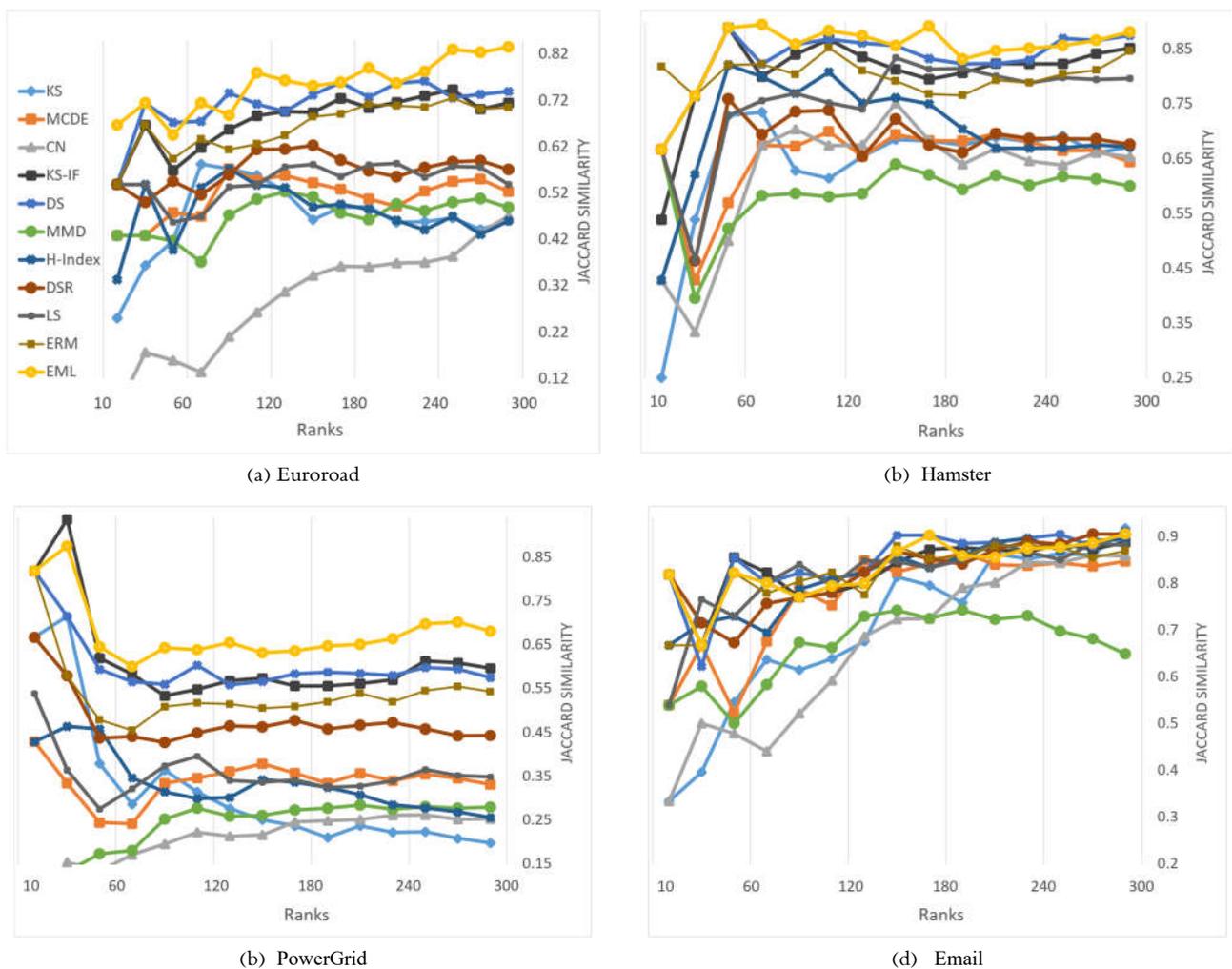

(a) Euroroad     (b) Hamster

(b) PowerGrid    (d) Email

Figure 3. Jaccard Similarity between the set of top-k nodes identified by different algorithms and the top-k nodes of the ground-truth.

We have done repetitive experiments over various networks to see how the Jaccard similarity will change if we change the value of $k$. Figure 3 shows the results of these experiments for four netwoks, Euroroad, Hamster, PowerGrid and Email. The Jaccard similarity by its nature is very sensitive to any error in R (i.e., missing a top influential node and not having it in the list). Therefore, even a score 0.7 shows a high degree of similarity between the two lists. As the results clearly depict, EML receives the highest Jaccard similarity score for the top 10 ranks over three datasets, Euroroad, PowerGrid, Email. On Hamster dataset, it gets the second highest JS score after the ERM model. This implies that EML outperforms the other models in identifying the most influential nodes in the networks. As $k$ increases, the performance of EML improves while staying at the top in three datasets, and in the other dataset declines, but stays at the top. This highlights that identifying the most influential nodes is more challenging for all the models, including EML, and there are rooms for improvement here.

Similar to previous section, the dynamics-sensitive model, DS, stays the second-best model and follows EML closely. These models are then closely followed by the KS-IF model, similar to what happened in the results of Figure 2. The agreement between the top models in the results of Figures 2 and 3 show that although the correlation of rankings results was showing the performance of the models overall, the Jaccard similarity results creates a more detailed picture of how the models perform in identifying the influential nodes at different ranks.

### 3.4 The Uniqueness of Ranking

The quality of an influential node ranking is not solely measured by the position of the nodes in the ranking, but also by the diversity of ranks in it. In fact, in a real network with thousands of nodes, where nodes have different structural and influential properties, it is unlikely that many nodes get the same influential ranking. In an ideal scenario, when the influential properties of nodes are different, they should be assigned with different unique ranks.

The common metric for measuring the uniqueness of ranking, that is also used by many other works [17, 31, 32], is Monotinicity Relation ($M(R)$), which gets values between 0 and 1, with a bigger value showing higher degrees of uniqueness in the ranking list, $R$. This metric is mathematically defined as follows:

$$M(R) = 1 - \left(\frac{\sum_{r \in R} n_r \times (n_r - 1)}{n_u \times (n_u - 1)}\right)^2$$

where $n_u$ is used to denote the number unique ranks in the Ranking list $R$, and $n_r$ being the number of nodes with the same rank $r$. Table 3 reports the monotonicity relation results for every model over different networks, where the top performance is highlighted by bold text. Across all datasets, EML achieves state-of-the-art performance outperforming other models (in three cases), or sharing the top place with another model (in the other five cases). In five out of eight datasets, EML achieves either absolute uniqueness (Delaunay dataset), or almost absolute uniqueness (PowerGrid, Email, PGP, Collaborations). For the other three datasets, the degree on uniqueness is very high as well. The lowest performance of EML in terms of uniqueness of ranking is on Citeseer dataset. It appears from the results that other models also found this dataset challenging to obtain a unique ranking. One can attribute the relatively low performance of models on this dataset to structural properties of nodes in this network. The degree distribution of this network follows a power-law distribution, with an average degree of 2.77, and a max degree of 99. This implies that there are many nodes in this network with similar degree and structural properties that may or may not have the same influential properties. Therefore, to the extent that a model relies on structural properties which are same among these nodes, that model fails to distinguish between the influentiality of the nodes and receive a low score for the uniqueness of ranking.

| Datasets | KS | MCDE | CN | KS-IF | DS | MMD | H-Index | DSR | LS | ERM | EML |
|---|---|---|---|---|---|---|---|---|---|---|---|
| Euroroad | 0.2430 | 0.6852 | 0.7310 | 0.938 | 0.7355 | 0.4442 | 0.2534 | 0.9921 | 0.7446 | **0.9978** | **0.9979** |
| Email | 0.8344 | 0.9437 | 0.9397 | 0.9992 | 0.9125 | 0.8954 | 0.8583 | **0.9999** | 0.9997 | 0.9978 | **0.9998** |
| Hamster | 0.8779 | 0.9544 | 0.9697 | 0.9853 | 0.4826 | 0.8895 | 0.8835 | 0.9544 | 0.9826 | 0.9848 | **0.9860** |
| Citeseer | 0.4346 | 0.6613 | 0.7940 | 0.9278 | 0.7733 | 0.5631 | 0.4452 | 0.9219 | 0.8376 | 0.9174 | **0.9500** |
| Collaborations | 0.7146 | 0.8963 | 0.9185 | 0.9962 | 0.1174 | 0.7915 | 0.7302 | 0.9994 | 0.9988 | 0.9909 | **0.9994** |
| PowerGrid | 0.3614 | 0.7566 | 0.8217 | 0.9678 | 0.7302 | 0.5927 | 0.3930 | 0.9994 | 0.8941 | **0.9999** | **0.9998** |
| Delaunay | 0.1330 | 0.8675 | 0.8848 | 0.9915 | 0.0580 | 0.6323 | 0.2252 | 0.9999 | 1 | 0.9695 | **1** |
| PGP | 0.5093 | 0.6856 | 0.793 | 0.9859 | 0.4312 | 0.6193 | 0.5172 | 0.9994 | 0.9743 | **0.9997** | **0.9997** |

Table 3. The uniqueness of ranking measure, M(R), for the rankings generated by the proposed model and the baselines.

The results in Table 3 also highlight the significant drop in performance for DS and KS-IF; two methods which achieved the second and third highest performances in the last two sections respectively. Among these two methods, DS performed the poorest, recording the lowest $M(R)$ among all models over four different datasets (Hamster, Collaborations, Delaunay, PGP). This means that many nodes in the ranking generated by DS have the same ranks. This is probably due to the fact that DS uses matrix multiplication for calculating nodes vitalities. In such calculation, the adjacency vector is the primary representation used for each node, and it will be highly correlated with the calculated vitality. Therefore, nodes with similar adjacency vectors will get similar influentiality, and eventually similar rank.

To visualize the actual difference between the rankings with absolute uniqueness, such as the rankings generated by EML, and rankings with low $M(R)$ score, we plot how the nodes are distributed among the ranks by plotting the number of nodes between consecutive ranks. We call this graph distribution of nodes at ranks. Figure 4 shows the distribution of nodes at ranks for four datasets.

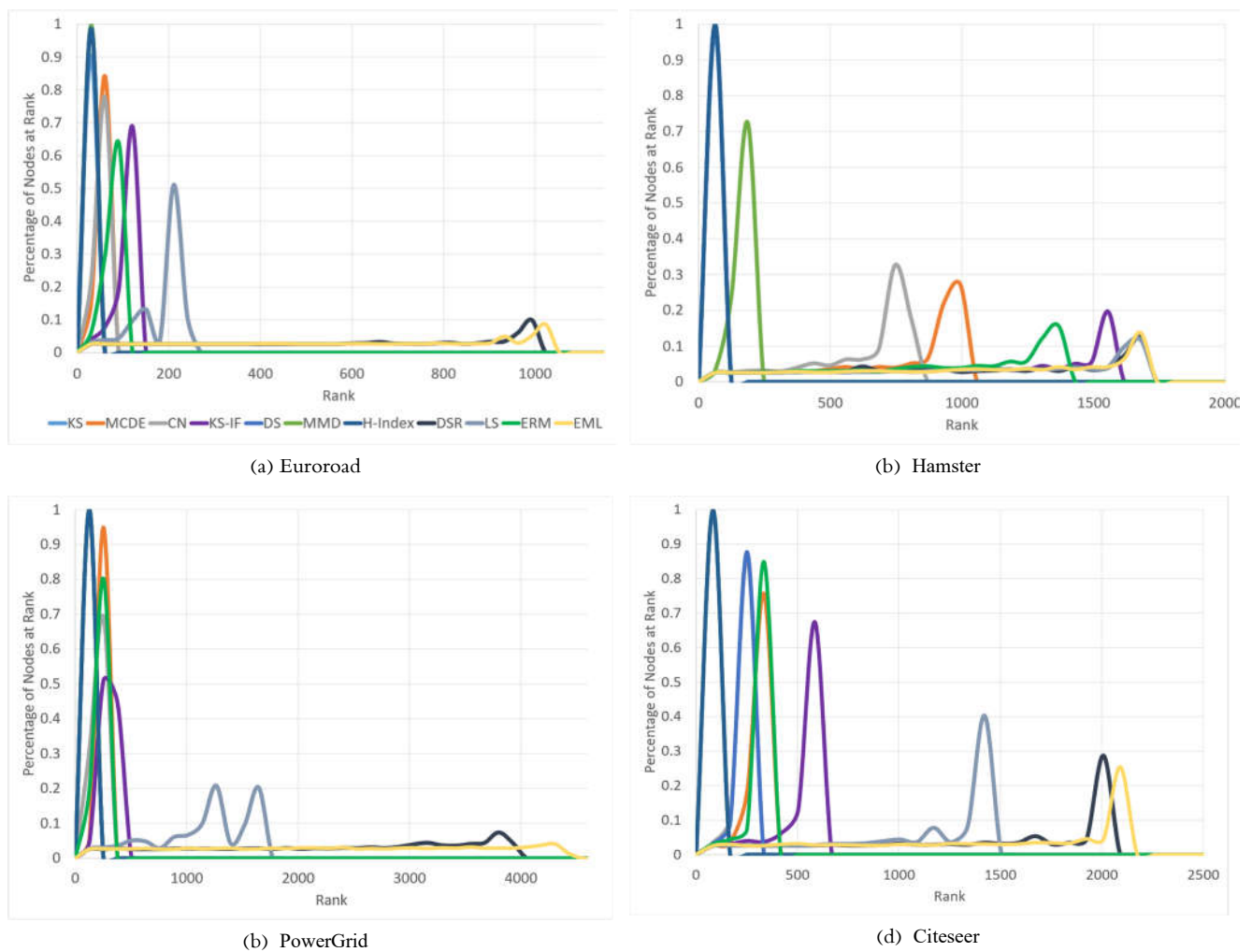

Figure 4. Distribution of rankings generated by different algorithms for four datasets. A uniform distribution shows indicates that the number of ranks is equal to the number of nodes, while a spiking distribution shows that many nodes have the same rank.

For PowerGrid dataset where the $M(R)$ score of the EML method show an almost absolute uniqueness (an $M(R)$ score of 0.9998), we can see that the distribution of ranks is uniform, meaning that there are equal number of nodes between ranks 20 to 30 same as the number of nodes between ranks 1030 and 1040, and same as the number of nodes between ranks 3570 and 3580. EML shows a similar performance across the two other datasets (Euroraod and Hamster), with assigning a slightly greater number of nodes to highest ranks. For the Citeseer dataset where EML records its lowest $M(R)$ score, we can see that the percentage of nodes at the highest ranks are greater than the other datasets, meaning that there are more nodes with the same high ranks in this dataset than to the other three datasets.

In contrast to EML, the distribution of nodes at ranks for other methods except DSR is highly non-uniform. The other methods are move biased to assign low ranks to nodes. For instance, the distribution of nodes at ranks for methods like DS, MCDE and MMD are spike-shape where the centre of the spike is at very low ranks. In two datasets (Euroroad and PowerGrid), most of the methods do not assign any rank higher than 400 to any node, leaving at least 50% of the nodes with similar ranks. Moreover, we see methods like KS, KS-IF and ERM that are based on shell-decomposition show good performance on a dense network like Hamster, while performing poorly on Euroroad and PowerGrid. One can attribute the difference in performance of these models to the number of shells in a network. When the network is dense, the number of shells in the network are higher, and the chance having two different nodes with the same shell level is low. However, in networks with low average degree, such as Euroroad and PowerGrid, the number of shells will be smaller, leading to a higher chance of having nodes with the same shell level, and the same influentiality ranking eventually.

**3.4 Practical Running Time**

The running time of algorithms in practice is the other criterion used in this study for comparing the performance of different methods. It is an indicator of scalability of models and can show practical useability of models. Figure 5 shows practical running time of the proposed method and the baselines on all datasets. For the proposed method, EML, practical running time refers to the training and prediction time. The results show that compared to other top performing models, namely, DS, DSR and ERM, EML has a lower practical running time on seven datasets, while it outperforms DS and DSR on PGP too, but has a slightly higher running time than ERM on the PGP. For the dynamics sensitive model, DS, the high running time was expected as it uses matrix multiplication for approximating the SIR dynamics. However, the gap between the running time of EML and DS on big graphs such as Delaunay and PGP are huge (running time of DS on these is graphs is 30 and 8 times bigger than that of EML). There is also a big gap between the running time of DSR and EML on big graphs. The high running time limits the applicability of these methods.

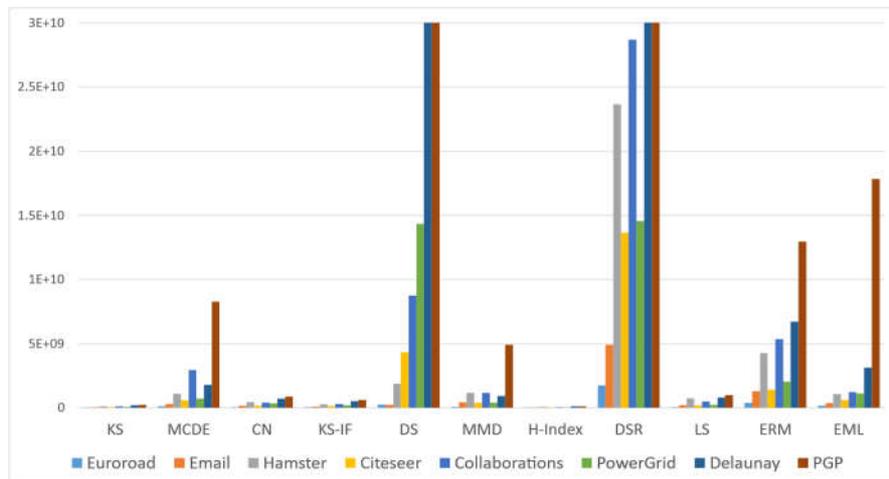

Figure 5. Practical running time of the models on all datasets.

Ultimately, it is evident that some of the methods studied in this research perform good with respect to some of the performance metrics, while performing poorly with respect to others. However, our proposed method, EML, was the only method that maintained a competitive performance across all the different performance metrics studied in this work. The main component of the EML model is the Support Vector Regression (SVR) machine installed at the heart of the model. We chose SVR because it can be easily trained with small training data, as well as its low training time, and its stable performance across different datasets. Table 4 shows the correlation of rankings with ground-truth for the EML method when it is trained with alternative machine learning models. To see the impact of the machine learning model on the performance of EML, we plugged 7 other machine learning models to the core of EML. The models include Multi-Layer Perceptron (MLP) [52], K-Nearest Neighbour (KNN) [53], Decision Tree (DT) [54], Random Forest (RF) [55], Adaboost (ADA) [56], Gradient Boosting Machines (GB) [57], and Voting Regressor (Voting) [58]. For each model, we ran Grid Search [59] to find the best set of hyper-parameters. For generating the results of Table 4, the optimal configuration for each model is used. As it can be seen, the SVR model with RBF kernel shows the highest stability in performance among all the models. In fact, over 4 datasets (Euroroad, Email, Hamster, Citeseer), SVR achieves the best performance, while in the other four datasets, it is just marginally outperformed by another model. No other model is showing the same stability as of SVR, meaning that other models such as ADA may perform well on few datasets (Collaborations, and PGP), but their performance on other datasets (Email, Hamster, Euroroad) is way off the top performance.

| Datasets | MLP | KNN | DT | RF | ADA | GB | SVR | Voting |
|---|---|---|---|---|---|---|---|---|
| Euroroad | -0.7478 | 0.8812 | 0.8807 | 0.8848 | 0.8784 | 0.8802 | **0.8905** | 0.8931 |
| Email | -0.4104 | 0.8998 | 0.8695 | 0.8958 | 0.7556 | 0.8481 | **0.9021** | 0.8851 |
| Hamster | 0.3201 | 0.8119 | 0.8 | 0.8608 | 0.8645 | 0.8409 | **0.8928** | 0.8742 |
| Citeseer | 0.8395 | 0.8117 | 0.8056 | 0.8052 | 0.8141 | 0.8072 | **0.8253** | 0.8171 |
| Collaborations | 0.2761 | 0.7863 | 0.7825 | 0.7699 | **0.787** | 0.7761 | 0.7701 | 0.7761 |
| PowerGrid | -0.5417 | 0.8408 | 0.8357 | 0.8412 | 0.8414 | 0.842 | 0.8416 | **0.8557** |
| Delaunay | 0.2629 | 0.7813 | 0.7806 | **0.7823** | 0.7819 | 0.7806 | 0.78 | 0.7777 |
| PGP | 0.2053 | 0.71 | 0.7003 | 0.6964 | **0.7324** | 0.721 | 0.7304 | 0.7311 |

Table 5. Correlation of rankings with ground truth for the EML method when trained with different machine learning models.

## 4. Conclusion

Many aspects of the modern human life are now impacted by how people are connected to each other, and how one individual can influence the others. By modelling human and social relationships as complex networks, problems such as vital node identification has attracted much attention within the research community. While most of the existing approaches of vital node identification rely on heuristic algorithms that are customized to specific scenarios, there is a need for more generalized and data-driven approaches. In this research, we propose a novel machine learning based approach, EML, for vital node identification. The proposed model uses the ground-truth vitality of a small portion of the nodes, 0.5% of the entire network, for training a support vector regression (SVR) model. The predictions of the trained SVR are then used in another function to predict the vitality of every node in the network. The EML model with a generic and data-driven design outperforms existing heuristic and case-specific models across different performance metrics such as correlation of rankings with ground-truth, ability to find the top-k most influential nodes, and uniqueness of ranking. Moreover, the model shows stability in its performance even with the parameters of the underlying dynamic changing. When compared with another state-of-the-art dynamics-sensitive approach, DS, EML not only shows a better sensitivity to the dynamic but is also capable of assigning unique ranks to the nodes, while DS fails to satisfy uniqueness of ranking. Among a number of state-of-the art models studied in this research, EML is the only model which its performance stays at top with respect to different criteria, while the other models perform well at some experiments, and fail at others. However, there is still a lot of room for improving EML. Future works can study the impact of changing the representation of the nodes and include other structural or dynamical features in the final feature vector. One can also study the performance of EML on other dynamics such as TIP model, SI, or linear threshold, or on other types of the networks such as temporal graphs or heterogenous networks.